\documentclass[twoside,12pt]{article}
\usepackage[]{lineno}
\usepackage{epsfig}
\begin{document}


%
%
\noindent
\vspace{0.5cm} \\
LHeC-Note-2013-001 GEN ~~~~~~~~~~\\
Geneva, \today \\
\begin{figure}[h]
\vspace{-1.6cm}
\hspace{14.9cm}
\includegraphics[clip=,width=.13\textwidth]{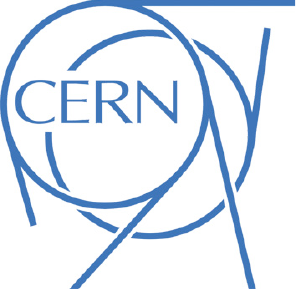}
\end{figure}
\begin{figure}[h]
\vspace{-3.0cm}
\hspace{5.5cm}
\includegraphics[clip=,width=0.35\textwidth]{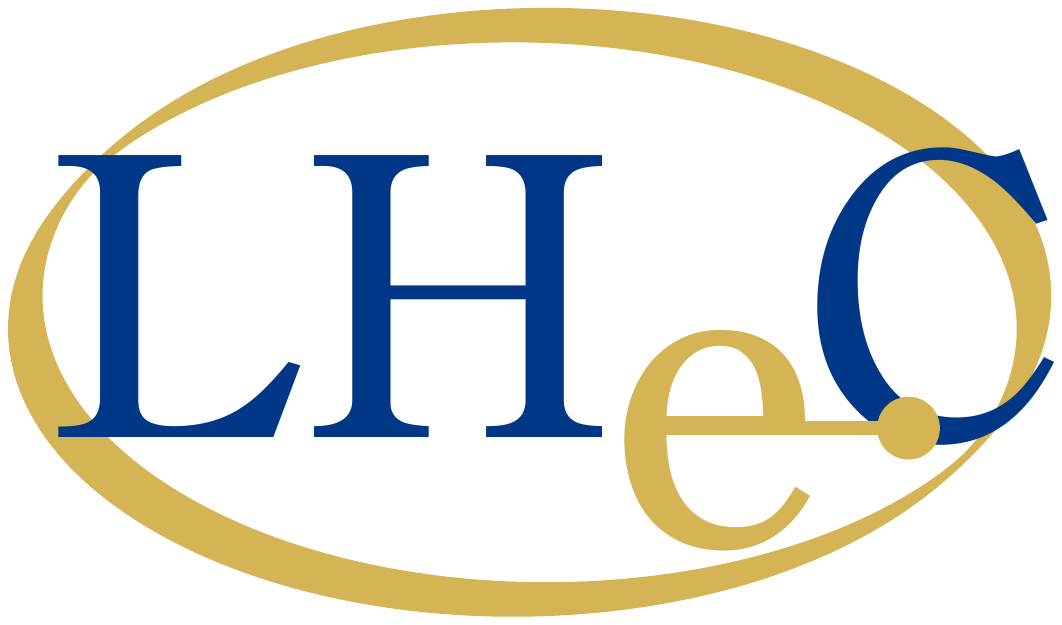}
\end{figure}
\begin{center}
\vspace{-1.6cm}
\begin{LARGE}
\bf{The Large Hadron Electron Collider} \\
\end{LARGE}
\vspace{0.5cm}
\begin{LARGE}
 \end{LARGE}
 \vspace{0.5cm}
%
O.\ Br\"{u}ning$^{1}$, M.\ Klein$^{2}$\\ 
$^1$CERN, 1211 Geneva 23, Switzerland\\
$^2$University of Liverpool, Physics Department, L69 7ZE, Liverpool,
UK
%
\\
\vspace{0.5cm}
For the LHeC Collaboration \\
to appear in Modern Physics Letters A
\vspace{0.5cm}
\\
$\bf{Abstract}$ \\
\vspace{0.3cm}
\end{center}
An overview is given on key physics, detector and accelerator
aspects of the LHeC, including its further development,
with emphasis to its role as the cleanest microscope
of parton dynamics and a precision Higgs facility. 
\\
%
%
%
%
\section{Deep inelastic scattering}	
The scattering of leptons off protons has  lead to fundamental
insight and corresponding historic progress in particle physics. In 1955, with a
beam of $E_e=0.2$\,GeV electron energy, a finite proton radius of about $0.74$\,fm
was discovered. Using a higher energy beam, of  $E_e \simeq 10$\,GeV,
the measurement of the proton structure function $\nu W_2=F_2(x,Q^2)$
at fixed Bjorken $x$ as a function of the four-momentum transfer
squared $Q^2$, performed  by the famous SLAC-MIT experiment, established the 
existence of partons as the smallest constituents of protons~\cite{Bloom:1969kc}. 
Ten years later, in 1978, a measurement of the polarization asymmetry
in $ep$ scattering at very low $Q^2$ determined the right-handed weak isospin charge of
the electron to be zero~\cite{Prescott:1978tm}, which was crucial for the
identification of the Glashow-Weinberg-Salam theory as the appropriate
description of the electroweak interaction. The first electron-proton 
collider, HERA, was built at DESY in eight years between 1984 and 1992. 
It extended the $Q^2$ range up to a few times $10^4$\,GeV$^2$
and explored the region of very low $x=Q^2/sy \geq10^{-4}$, for
$s=4 E_e E_p \simeq 10^5$\,GeV$^2$ and the inelasticity  $y \leq 1$.
With HERA, deep inelastic
scattering (DIS) physics made enormous progress in the understanding of the
proton's structure, of the quark-gluon dynamics and its theoretical description
within Quantum Chromodynamics (QCD) and also in the search for new phenomena
beyond the Standard Model (SM) of particle physics~\cite{Klein:2008di}.
There would nowadays
be no quantitative description of LHC physics, and 
notably the Higgs production cross section, which at the LHC is
dominantly due to gluon-gluon ($gg$) fusion, would not be known
without the parton dynamics information deduced mainly from HERA.

The Large Hadron electron Collider (LHeC) is the next logical and a big step in the
evolution of DIS physics as part of the accelerator exploration of the energy frontier.
New phenomena in DIS may appear 
at high masses of new particles, as the Higgs or lepto-quarks, at very high
$Q^2$ exceeding the masses squared of the weak bosons $W$ and $Z$ and also
at very low $x \propto 1/s$ which at the LHeC extends down to $x \simeq 10^{-6}$. 
The LHeC kinematic range exceeds HERA's by a factor of about $20$, due to the combination of
a $7$\,TeV proton beam from the LHC and a new $60$\,GeV electron beam.
Its luminosity is projected to be as high as possibly $10^{34}$\,cm$^{-2}$s$^{-1}$,
with a default design value of $10^{33}$\,cm$^{-2}$s$^{-1}$. This is almost a thousand times higher than
HERA's luminosity, and it makes the LHeC a potential precision Higgs production
facility and enables a huge variety of new measurements and searches.

There was unfortunately no time given to operate HERA with deuterons nor heavy ions. 
Therefore the knowledge from lepton-nuclear scattering currently relies on fixed target
data only. The LHeC  extends the kinematic range with deep inelastic electron-ion
scattering by almost four orders of magnitude. A huge discovery potential there
appears in $eA$ regarding new phenomena in nuclear parton dynamics, nuclear PDFs
and the initial state of the Quark-Gluon Plasma (QGP). At lower energies 
concepts for electron-ion colliders are being developed also~\cite{newsletter}.

Basic LHeC design solutions
have recently been layed out in detail in a refereed conceptual design 
report (CDR) on the physics, accelerator
and detector concepts~\cite{AbelleiraFernandez:2012cc}. These have been 
summarised in~\cite{AbelleiraFernandez:2012ni}
and updated in \cite{AbelleiraFernandez:2012ty}  mainly in view of the 
Higgs discovery~\cite{:2012gk,:2012gu}.
The following paper presents the detector design concept, a few highlights of
the physics program and  summarizes the accelerator design
as well as sketching  directions for the future development of the LHeC.

%
%
\section{LHeC detector design}
The LHeC is the second electron-hadron collider following HERA.
Its physics programme demands a very high level of precision,
as for the measurement of  the strong
coupling constant $\alpha_s$ to per mille uncertainty, 
and it requires  the reconstruction of complex final states, 
as appear in  charged current
Higgs production and decay into $b \overline{b}$ final states.
As a consequence of the asymmetric electron and proton beam
energy configuration, 
the detector acceptance has to extend as close as possible to
the beam axis. The dimensions of the
detector are constrained by the radial extension
of the beam pipe, elliptic due to synchrotron radiation,
in combination with a polar angle coverage extending  down 
to about $1^{\circ}$ and $179^{\circ}$ for forward going final
state particles and backward scattered electrons at low $Q^2$,
respectively. 
\begin{figure}[th]
\begin{center}
\centerline{\epsfig{file=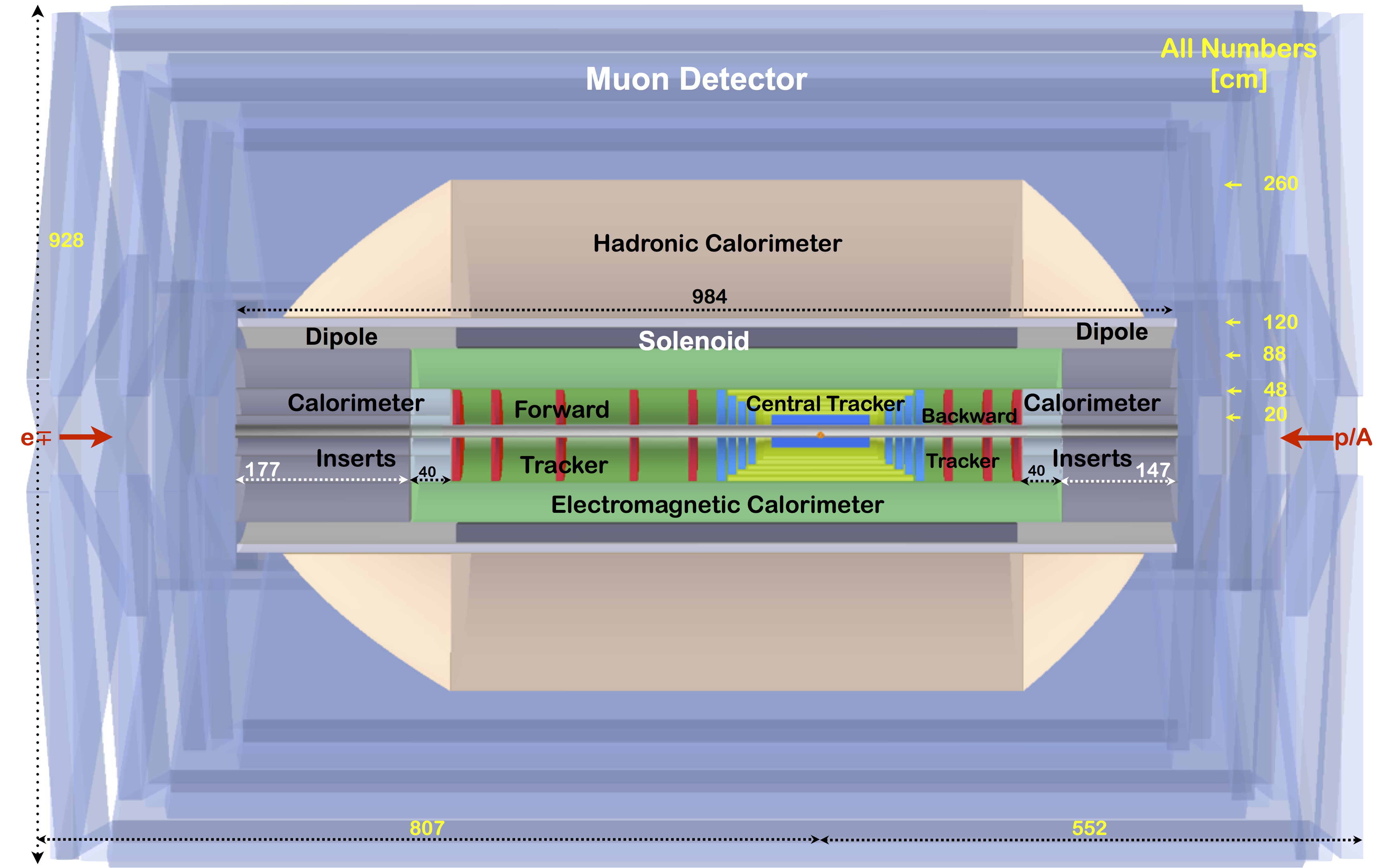,scale=0.5}}
\end{center}
\caption{\footnotesize{
An $rz$ cross section of  the LHeC detector
in its baseline design with 
the solenoid and dipole magnets  placed between the electromagnetic and the hadronic
calorimeters. The interaction point   is surrounded by
a central tracker system, complemented by large forward and backward tracker
telescopes, and followed by sets of calorimeters, see text.
The detector dimensions are $\approx13.6$\,m longitudinally to the beam
and  $\approx9.3$\,m in diameter, which may be compared with the
CMS dimensions of $21 \times 15$\,m$^2$. }
\label{figdet}}
\end{figure}

A cross section of the central, baseline detector
is given in Fig.\,\ref{figdet}.
 In the central barrel, the following
detector components are currently considered:
a central silicon pixel detector surrounded by
silicon tracking detectors of strip or possibly strixel technology;
an electromagnetic LAr calorimeter inside
 a $3.5$\,T solenoid and
a dipole magnet, with a $0.3$\,T field on axis,
 required to achieve head-on collisions; 
a hadronic tile calorimeter  serving also 
for the solenoid flux return
and a muon detector, so far for muon identification only,
relying on the precise inner tracking for momentum measurements.
The electron at low $Q^2$ is scattered into the
backward silicon tracker  and its energy is measured
in backward calorimeters. In the forward region,
components are placed for tracking and for
calorimetry to precisely reconstruct jets 
over a wide energy range up to O(TeV).

Simulations of tracking and calorimeter performance
were used to verify the design, while a full detector
simulation is not yet available. The momentum resolution
based on the central tracker is 
$\delta p_t/p_t^2 = 6 \cdot 10^{-4}$\,GeV$^{-1}$ which
translates to a radial impact parameter resolution of
$10$\,$\mu$m. Combined with the extension of
the beam spot of $~7$\,$\mu$m in both transverse directions
this promises to be a very precise heavy quark tagging
environment with the biggest challenge of a good forward direction
performance. The simulated resolution of the central
electromagnetic liquid argon calorimeter  (LAr) is 
$\sigma/E = 8.5/\sqrt{E/\rm{GeV}} \oplus 0.3 $\,\%. The hadronic energy
resolution, from a first combined LAr and scintillator tile calorimeter
simulation is
$\sigma/E = 32/\sqrt{E/\rm{GeV}} \oplus 8.6 $\,\%.

The CDR~\cite{AbelleiraFernandez:2012cc} also contains designs
for forward and backward tagging devices for diffractive
and neutron physics and for photo-production and
luminosity measurements, respectively.
The radiation level at the LHeC is much lower than in $pp$,
and the $ep$ cross section is low enough for the experiment
not to suffer from any
pile-up, which are the two most demanding constraints
for the ATLAS and CMS  detectors and their  upgrades for the HL-LHC. 
The choice of components for the LHeC detector  can  rely on the
experience obtained at HERA, at the LHC, including 
its detector upgrades currently being developed, and also on 
detector development studies for the ILC. 
The detector development, while requiring prototyping,
may therefore proceed without an extended R\&D program.

The time schedule of the LHeC project is given by 
the LHC and its upgrade project, which 
demand a detector to be ready within about $10-12$ years.
A first installation study was made  
considering pre-mounting
the detector at the surface, lowering  and installing it at IP2.
The detector is small enough to fit into the L3 magnet 
structure of $11.2$\,m diameter, which is still
resident in IP2 and would be available as mechanical support.
Based on the design, as detailed in the CDR, it is estimated
that the whole installation can be done in $30$\,months, 
which appears to be compliant with the operations 
currently foreseen in the LS3 shutdown in the early twenties.
%
%
\section{Physics with the LHeC}
\subsection{Overview}
With its unprecedented precision, deep
inelastic scattering range and resolution in probing partonic 
interactions, the LHeC has a huge scientific potential 
as has heen elucidated in~\cite{AbelleiraFernandez:2012cc}. By completely determining,
for the first time, the proton, neutron and nuclear parton densities, it  adds
considerably to the capabilities of the existing LHC experiments and
the HL-LHC upgrade program,
 for example in terms of
precision studies of Higgs properties, see below, 
and sensitive range in high mass LHC searches,
see~\cite{AbelleiraFernandez:2012ty}. 
Following Ref.~\cite{AbelleiraFernandez:2012ni} one may classify the physics
of the LHeC into six, partially overlapping research
categories: i) discoveries in QCD, Higgs, BSM and top
quark physics; ii) relations to the LHC; iii) gluon distribution and precision DIS;
iv) parton structure of nucleons and photons, perturbative QCD and
non-DGLAP evolution; v) heavy ion physics, including deuterons, and
modified parton distributions (GPDs, diffractive, unintegrated), and vi)
extension of HERA measurements as of the longitudinal 
structure function or vector mesons production.
For the current overview,  two  important and comprehensive 
subjects are selected here for a
more detailed presentation, the precision measurement of $\alpha_s$
and the potential for Higgs physics with the LHeC, both being related 
to the determination of the gluon distribution.

Every step into a new region of phase space and intensity can lead to new
observations as happened in DIS with the discoveries of scaling at SLAC or 
of the striking role of the gluon at HERA, the self-interaction of which
gives mass to the baryonic matter. DIS with the LHeC
may lead to discovering unexpected substructure phenomena, see for
example~\cite{Csaki:2012fh}, of the heaviest known particles,
the W,Z, top or even the Higgs to possess structure, 
or it may become crucial for disentangling contact interaction
phenomena which could be observed at the LHC with multi-TeV mass scales.
It may be discovered that there is $no$ saturation of the gluon density, despite common
belief, the odderon or instanton may eventually be found or, similarly,
the application of PDFs to describe LHC phenomena could become questionable
when factorization could be observed to not hold, not just in diffraction
but possibly in inclusive scattering too. Nature keeps holding surprises
which is the overriding reason for the LHeC to be built.
\subsection{The strong coupling constant and precision DIS}
Deep inelastic scattering is an ideal process for the determination
of the strong coupling constant, which determines the scaling
violations of the parton distributions. Theory is presently 
calculated to NNLO in perturbative QCD with elements already
available to N$^3$LO, see~\cite{Blumlein:2012bf}. Despite major efforts over
the past nearly $40$ years, since the discovery of asymptotic freedom,
and a plethora of $\alpha_s$ determinations, there is no accurate value
of $\alpha_s$ available~\cite{job1012}, with a precision comparable to the weak
coupling constant, and a number of severe problems remains to be solved.
Questions regard the (in)consistency 
of previous DIS data, the (in)consistency of inclusive DIS and jet based 
determinations, both in DIS and Drell-Yan scattering, or the
treatment leading to the world average on $\alpha_s$ and its 
uncertainty~\cite{Bethke:2011tr}.
 The LHeC has the potential to provide a new, coherent
data base, from neutral and charged current DIS 
including heavy quark
parton distribution measurements, with which 
an order of magnitude improved experimental determination
of $\alpha_s$ becomes possible. This is of crucial importance
for QCD, for predictions of LHC cross sections, notably that of
the Higgs production, discussed below, and for the 
predictions of grand unification of the electromagnetic, weak and
the strong interactions at the Planck scale. 
It is also long time to challenge the lattice QCD $\alpha_s$ results, which
seem to be most accurate but stand on different grounds 
than the classic data based measurements
exhibiting variations which are non-negligible~\cite{job1012}.

Two independent fit approaches have been undertaken
in order to verify the potential of the LHeC to determine $\alpha_s$.
These analyses used a complete simulation of the experimental
systematic errors of the NC and CC pseudo-data
and higher order QCD fit analysis techniques,
see the CDR~\cite{AbelleiraFernandez:2012cc} for details.
The total experimental uncertainty on $\alpha_s$ is estimated to be $0.2$\,\%
from the LHeC alone and $0.1$\,\% when combined with HERA. 
Relying  solely on inclusive DIS $ep$ data at high $Q^2$,
this determination is free of higher twist, hadronic and nuclear corrections,
unlike any of the recent global QCD fit analyses.
 There are known further, parametric, uncertainties in DIS determinations
of $\alpha_s$. These will be much reduced
with the LHeC as it resolves the full set of parton distributions,
$u_v,~d_v,~\overline{u},~\overline{d},~s,~\overline{s},~c,~b$
and $xg$ for the first time, 
providing $x$ and $Q^2$ dependent constraints not
``just" through the fit procedure. 
The LHeC therefore has a huge
power in the determination of PDFs which cannot be replaced nor
challenged by the yet important constraints inherent in
precision Drell-Yan data at the LHC~\footnote{This question is sometimes raised
in discussions but it is clear that the determination of $x$ and $Q^2$
in the DIS measurements, from both the scattered lepton and the hadron
over  a huge range of $~5$ orders of magnitude, combined with the theoretical
advantage of involving only one hadron is the appropriate way to measure PDFs.
This becomes immediately obvious from the comparison of HERA and Tevatron
results on PDFs, see also the discussion in~\cite{AbelleiraFernandez:2012ty}.}.
Recently a six-flavour variable number scheme has been 
proposed~\cite{Pascaud:2011zt}, 
in which it is predicted that the
top contribution to proton structure has an on-set much below the threshold of its production in a massless scheme. This may lead to the concept
of a top quark distribution which completed the
set of PDFs measurable with the LHeC.

Regarding the challenging precision on $\alpha_s$ one needs to not only
measure PDFs more accurately but control also the heavy 
quark theory and experimental input.
The measurement of the charm structure function in NC at the 
LHeC will
determine the charm mass parameter to $3$\,MeV, which is expected to
correspond to an $\alpha_s$ uncertainty well below $0.1$\,\%. 
Due to the huge range
in $Q^2$ and the high precision of the data,  decisive tests will
also become available for answering the question whether the
strong coupling constants determined with jets and in inclusive DIS
are the same. If confirmed, a joint inclusive and jet analysis has the 
potential to even further reduce the uncertainty of $\alpha_s$.

Matching this outstanding experimental precision requires
future LHeC based analyses on inclusive cross sections 
to be performed at N$^3$LO
pQCD for reducing the scale uncertainty. The ambition to
measure $\alpha_s$ to per mille accuracy thus represents
a vision for a renaissance of the theoretical and experimental
 physics of deep inelastic scattering
which is a major task and fascinating prospect of the LHeC enterprise.
\subsection{Higgs in electron-proton scattering}
In the Standard Model,  the Higgs field is responsible for generating masses 
of  the weak gauge bosons as well as the elementary fermions, in a mechanism
through absorption of Nambu-Goldstone bosons arising in spontaneous 
symmetry breaking. The simplest representation of the mechanism adds 
an extra field to the  theory which is a scalar, $J^{CP}=0^{++}$, that is 
referred to as the SM Higgs boson ($H$). Recent exciting developments
following the discovery of a new boson of mass $M_H \simeq 125$\,GeV
by ATLAS~\cite{:2012gk} and CMS~\cite{:2012gu} indicate that
this particle most likely indeed is the Higgs boson. The exploration of 
its properties has begun to become a focus of modern particle physics.
The observed  Higgs mass value leads to a rather
large number of decay modes which will enable detailed investigations
of the properties of that boson to be made. At the LHC,
background, theoretical and experimental conditions, as large
pile-up, make it not easy to achieve
high precision Higgs related measurements.
Therefore various lepton-lepton and photon-photon 
collider configurations have been vigorously studied,
while the genuine prospects of Higgs physics at the LHC
are being investigated also, much related to ATLAS and CMS
detector upgrade designs. 

At the LHC, the Higgs is dominantly produced
 via a top loop in  $gg \rightarrow H$ fusion. A smaller
 fraction stems from the  fusion of vector bosons ($V=W,Z$)
into Higgs.
There is also the associated $VH$ production mechanism in $pp$.
The interest in Higgs physics with the LHeC primarily comes from its
clean production mechanism, based on $H$ emission from 
 $W$ or $Z$ $t$-channel exchange in CC or NC scattering,
 and low QCD backgrounds.  The cross section of
 the SM Higgs production in  polarized $e^-p$ CC 
 scattering~\footnote{
 The cross section in $e^+p$ is lower, about $60$\,fb, due to the
 involvement of down instead of up quarks. Since very high
 luminosities in the linac-ring configuration will be limited to electrons
 and a high degree of polarization for positrons is unlikely achievable either,
 the $e^+p$ configuration is inferior to $e^-p$ for LHeC Higgs physics. In NC
 the cross section is about $20$\,fb.} 
 is about  $200$\,fb at the default LHeC energies
  of $E_e=60$\,GeV and $E_p=7$\,TeV. 
 Therefore the $e^-p$ $H$ production cross section is 
 about as large as the  $Z$-Higgsstrahlung 
  cross section at
 an $e^+e^-$ collider above $H+Z$ threshold energies.
  Compared to the LHC, the
 $ep$ configuration has the advantage of a cleaner final
 state reconstruction due to the presence of only one hadronic vertex
 and the absence of pile-up. It is important also that the
 theoretical uncertainties of Higgs production in $ep$ are 
 small~\cite{Blumlein:1992eh}. The LHeC thus appears to be a
 very attractive facility for Higgs physics complementing the LHC and
 an $e^+e^-$ machine.
\begin{figure}[htbp]
\centerline{
\epsfig{file=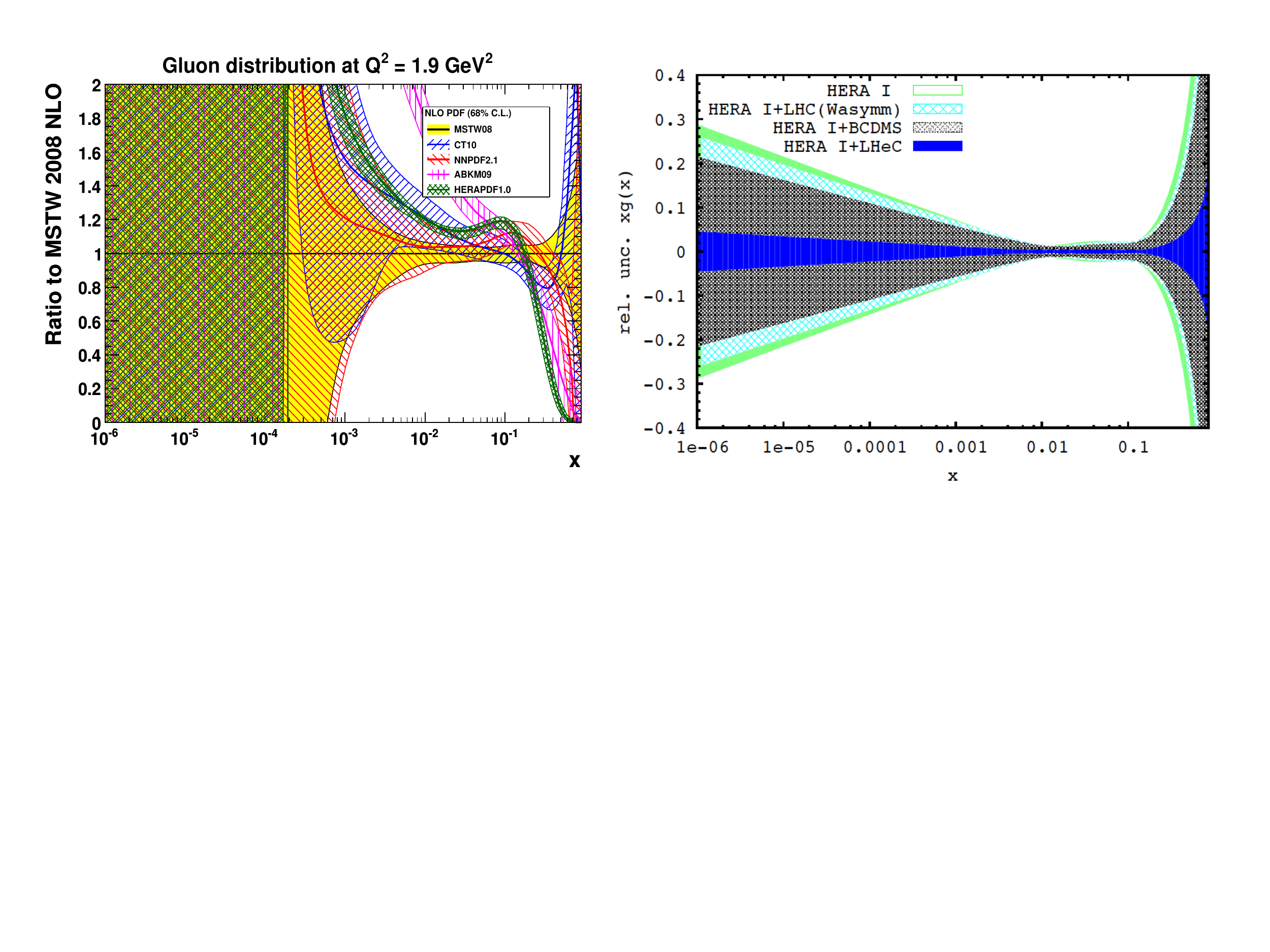,scale=0.74}}
\vspace{-6.9cm}
\caption{\footnotesize{Uncertainty of the gluon distribution at
$Q^2 = 1.9$\,GeV$^2$ as a function of Bjorken $x$. 
Left: Recent gluon distribution determinations and their uncertainties, 
plotted as a ratio to MSTW08. Below $x \simeq 10^{-3}$
the HERA data have vanishing constraining power due to
kinematic range limitations and the gluon is not determined
at low $x$. It is for the LHeC to discover whether 
$xg$ saturates or not and whether indeed the DGLAP
equations need to be replaced by non-linear parton 
evolution equations as BFKL$^{16}$.  
 At large $x \geq 0.3$ the gluon distribution 
becomes very small and large variations appear in its
determination, differing by orders of magnitude, which 
is related to uncertainties of jet data, theory uncertainties
and the fact that HERA had not enough luminosity to
cover the high $x$ region where, moreover,  the sensitivity
to $xg$ diminishes, as the valence quark evolution
is insensitive to it.
The situation can be expected to improve with
LHC jet and possibly top$^{17}$ and the HERA II data. 
Right: Experimental uncertainty of $xg$ based on HERA alone
and in various combinations, see the CDR$^5$. 
At large values of e.g. $x=0.6$ the LHeC can be expected to determine
$xg$ to $5-10$\,\% precision (inner blue band). At small $x$ a few per cent precision
becomes possible, compare right with left.
Note that
the non-LHeC low $x$ uncertainty bands  below $x \simeq 10^{-3}$
(right) remain narrow solely as an artifact due to the parameterization
of xg.}}
   \label{figxg}
\end{figure}

 Prior to the Higgs discovery, for the LHeC design a study was made of
 the prospect to reconstruct the decay of $H \rightarrow b \overline{b}$
 as this dominates, to $60$\,\% , the branching fractions
but is very difficult to precisely measure at the LHC for QCD background
resons.
 The result~\cite{AbelleiraFernandez:2012cc,AbelleiraFernandez:2012ty},
from an initial cut based analysis, is a signal-to-background ratio of $1$
and a statistics which allows to determine the this cross section 
to $3$\,\% statistical accuracy with $100$\,fb$^{-1}$ of luminosity. 
This result has a number of implications: i) it demonstrates that the
$ep$ collider has a huge potential for precision Higgs physics; ii) 
with luminosities of order $10^{34}$\,cm$^{-2}$s$^{-1}$ 
complementary access becomes 
possible to further decay modes such as into fermions
$H\rightarrow \tau\tau, c\overline{c}$, both challenging at the LHC, $c$
involving a second generation coupling,
and also the $H$ decay into bosons,  $WW,~ZZ,~\gamma\gamma$,
from a clean $WW$ initial state, the former delivering a potentially
clean measurement of the $H$ to $WW$ coupling;
iii) with the specific $ep$ configuration unique measurements
of the CP properties are in reach with access to CP odd admixtures and/or 
precision measurements of the CP even (SM) 
eigenvalues~\cite{Biswal:2012mp}. 
\\
\vspace{-2cm}
\begin{figure}[hbt]
\centerline{\epsfig{file=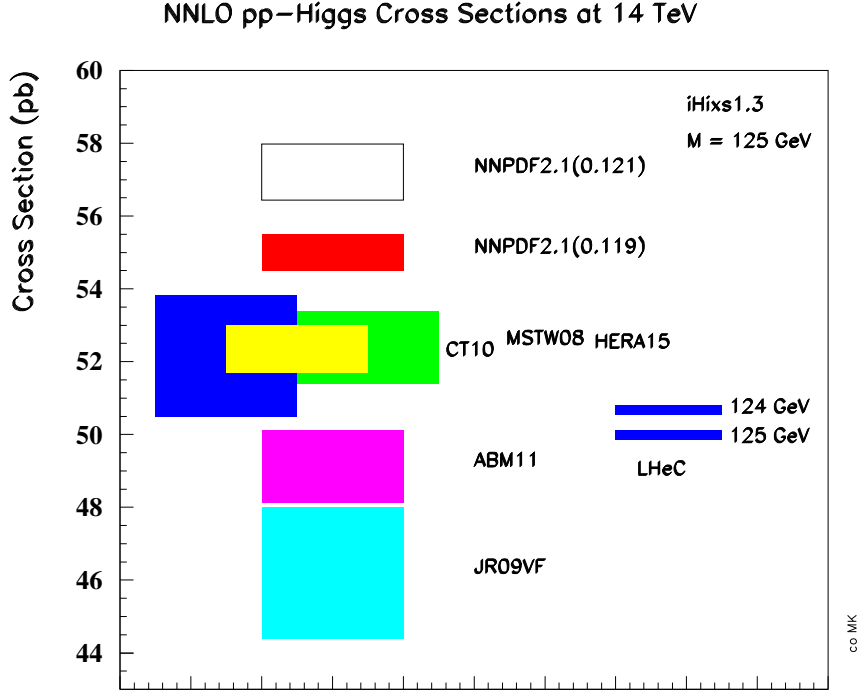, scale=0.6}}
\vspace{-4cm}
\caption{\footnotesize{NNLO calculation of the Higgs  production 
cross section in $pp$ scattering at the design LHC energy
using the iHixs program. The cross section
is calculated at a scale of $M_H/2$. The bands on the left
side represent the uncertainties of the various PDF sets
available to NNLO as marked. The PDF4LHC convention 
excludes ABM11, JR09VF, HERA and extreme values of $\alpha_s$
arriving in this calculation to roughly  $5$\,\% uncertainty from
PDF variations to which one would add an about $10$\,\%
from scale uncertainty, as this picture looks different when
$M_H$ is used, see text, and about $5$\,\% due to $\alpha_s$. 
The full experimental uncertainty estimated with the LHeC
PDFs, as detailed in the CDR and plotted at the right
column, is about $0.3$\,\%, with 
a similar uncertainty to be added from $\alpha_s$
discussed above. From these
two sources therefore, the LHeC would provide the means
to derive Higgs mass values from LHC cross section measurements.
}}
   \label{figH}
\end{figure}

At the LHeC one probes new physics
at the cleanly separated
$WWH$ and $ZZH$ 
vertices with a simpler final state, no pile-up and
knowing the directions of the struck quark.
Measurements of couplings have to be precise as, for example,
the $H$ to $WW$ and $ZZ$ couplings, when measured with better than $8$\,\%
accuracy, could allow accessing a composite Higgs structure~\cite{Gupta:2012mi}.
The prospects for Higgs physics with the LHeC are remarkable 
and deserve to be studied deeper.

A salient further aspect of $ep$ assisting to make the LHC a precision
Higgs physics facility is the superb measurement of the 
PDFs and $\alpha_s$ in $ep$ with the LHeC. 
The dominant production mode for the Higgs
in $pp$ is $gg$ fusion and therefore the cross section is proportional 
to the product of $\alpha_s$ and $xg$ squared. The LHeC leads to 
a much improved determination of the gluon density over $5$ orders
of magnitude in Bjorken $x$, extending to large $x$ as is illustrated 
in Fig.\,\ref{figxg}. This is at the origin of a huge improvement of
the knowledge, based on pseudo LHeC data, of the Higgs production cross section
at the LHC, shown in Fig.\,\ref{figH} and calculated 
with $iHixs$~\cite{Anastasiou:2012hx},
in comparison with the available 
NNLO PDF determinations. It thus is possible to essentially remove or control
the theory uncertainties of $H$ measurements at the LHC which now
are of the order of $10$\,\%, depending on PDF assumptions and the
admixture of $VH$ events in $H$ data samples. Naturally this will lead
to the requirement of N$^3$LO cross section calculations combined with
most precise $\alpha_s$ and N$^3$LO PDF determinations as can
emerge in a decade hence with the LHeC and intense theoretical 
developments~\footnote{The scale dependence of the $gg \rightarrow H$
production cross section at NNLO is still large.
The choice of scales in  pQCD calculations
is to some extent arbitrary but indicative of missing higher order terms.
For $M_H=125$\,GeV, at $14$\,TeV cms energy
using $iHixs$, one obtains a cross section of $52.5$\,pb with an
uncertainty of $(+1,-1.6)$\,\% for the MSTW08 PDF set using 
$\mu_r=\mu_f=M_H/2$ (1/2,1/2),
which gives the yellow band in Fig.\,\ref{figH}. If instead one sets, as
mostly is done, $\mu_r=\mu_f=M_H$ (1,1), the cross section is $47.9$\,pb,
i.e. reduced by $9$\,\%. This is mainly due to the renormalization
scale dependence as is seen by independent
variations of $\mu_r$ and $\mu_f$. The result is $53.1$\,pb and $47.2$\,pb for
the cases (1/2,1) and (1,1/2), respectively.}. 
The ILC in this context would provide a measurement of the
width and precision data which delivered  further insight even though
the challenge to reach $10^{34}$\,cm$^{-2}$s$^{-1}$
luminosities with positrons at the
linear collider is considerable and the effort immense. Higgs physics can be done
best with the combination of $pp$, $ep$ and $e^+e^-$ colliders.
It may lead beyond the Standard Model and the SM Higgs,
which could be composite~\cite{Kaplan:1983fs}.
%
%
\section{Accelerator design}
\subsection{LHeC project}
\begin{figure}[htbp]
\centerline{\epsfig{file=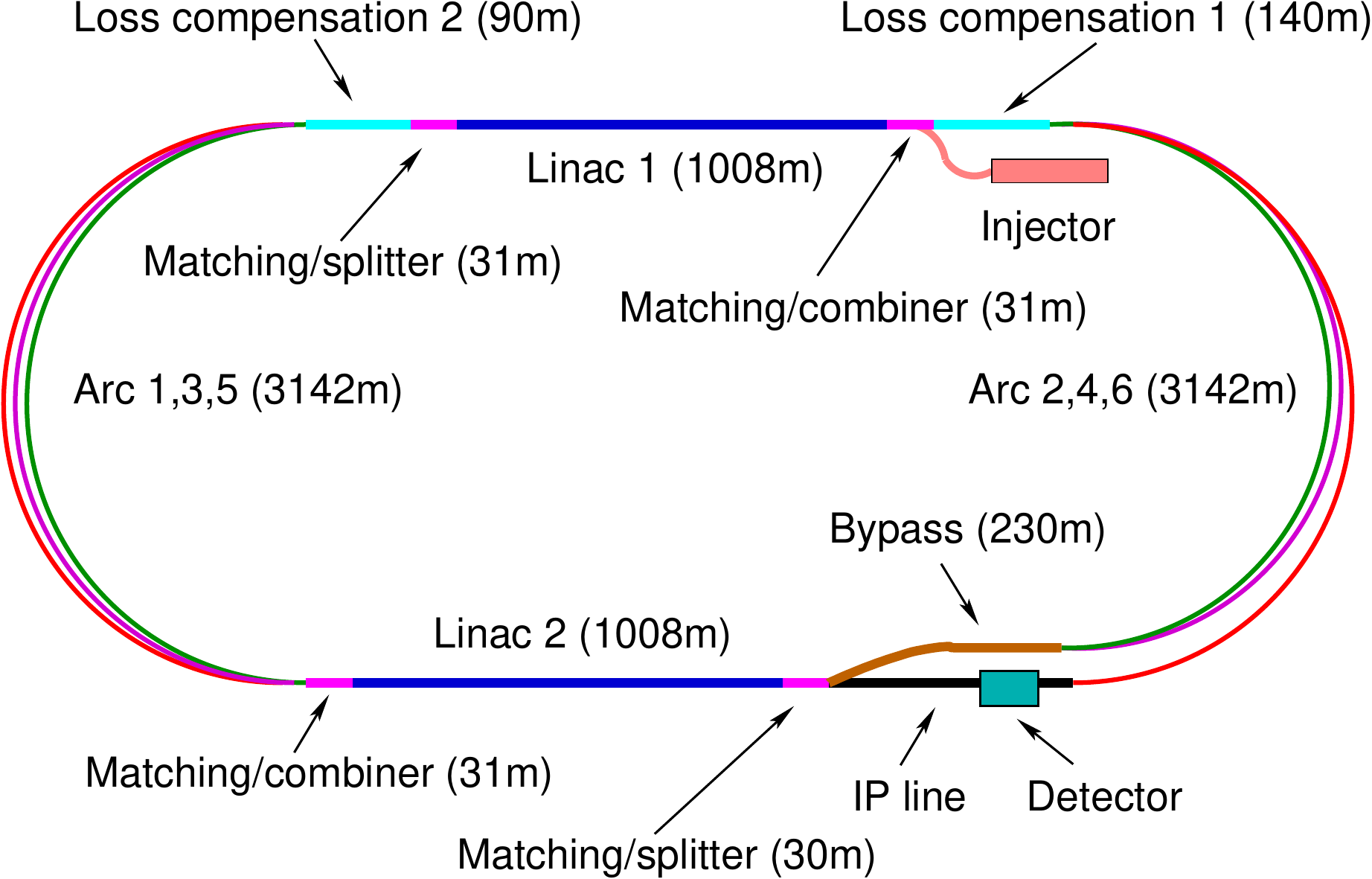,scale=0.7}}
\vspace*{0.2cm}
\caption{\footnotesize{
Schematic view of the default LHeC racetrack configuration.
Each linac accelerates the beam to $10$\,GeV, which leads to a
$60$\,GeV electron energy at the interaction point after three
passes through the opposite linear structures of $60$ cavity-cryo modules
each. The arc radius is about $1$\,km, mainly determined
by the synchrotron radiation loss of the $60$\,GeV beam which 
is  decelerated for recovering the beam power
after having passed the IP.
The default tunnel circumference is $1/3$ that of the LHC.
The tunnel is designed to be tangential to IP2.
Detailed civil engineering considerations are described in the CDR.
}}
\label{figacc}
\end{figure}

The LHeC is an electron-proton ($ep$) and electron-ion ($eA$)
complement of the Large
Hadron ($pp$ and $AA/p$) Collider, 
with which lepton-quark interactions can be explored at the TeV energy
scale.  In summer 2012 an extensive report, the CDR,
was published~\cite{AbelleiraFernandez:2012cc}, in which a new electron 
beam accelerator was designed, as a ring mounted on top of the
LHC ($RR$ option) and as a multiple pass, energy recovery linac in 
a racetrack configuration ($LR$ option), sketched in Fig.\,\ref{figacc}.
The LHeC is designed to run simultaneously with 
$pp$ (or $AA$) collisions.
LHeC operation is fully transparent to the other LHC experiments thanks to 
the low lepton bunch charge and resulting small beam-beam tune shift 
experienced by the protons.
After careful consideration of installation issues and parameters of
the electron beam, preference was given to the $LR$ option
for the next phase of design as this is rather independent of the LHC. 
Early considerations of linac-ring  electron-proton colliders were published
in~\cite{GrosseWiesmann:1988it,tigner}.

The LHeC design study has been pursued under the auspices 
of the European Committee for Future Accelerators
(ECFA), the Nuclear Physics European Collaboration Committee (NuPECC) 
following a CERN SPC mandate.
ECFA has released a supportive statement in November 2012, following 
the recommendations of an ECFA study group, while NuPECC, already in 
2010, decided to put the LHeC on the long range map for the future
of European nuclear physics. The combination of high energy 
electron-proton and electron-ion physics makes the LHeC an important
example which unites interests from accelerator based particle with 
nuclear physics. Following a mandate of CERN, a few years are now foreseen for
next developing key LHeC technologies in international 
collaborations, see below. 

\subsection{Luminosity prospects and power consumption}
The luminosity $L$ for the LHeC in its linac-ring configuration
is determined as
\begin{equation}
\label{LLR}
L = \frac{N_e N_p f \gamma_p}{4 \pi \epsilon_p \beta^*},
\end{equation}
where $N_e=10^9$ is the number of electrons per bunch, 
$N_p=1.7 \cdot 10^{11}$ the
number of protons per bunch, $f = 1/\Delta = 40$\,MHz the bunch frequency
with the bunch distance $\Delta=25$\,ns, $\epsilon_p=3.7$\,$\mu$m 
the normalized proton transverse beam emittance 
and $\beta^*=0.1$\,m the value of the
proton beta function at the IP, assumed to be equal in 
$x$ and $y$. The just quoted numbers are taken from the CDR.
They correspond to the nominal LHC proton beam parameters 
and lead to a peak luminosity of $10^{33}$\,cm$^{-2}$s$^{-1}$.
The electron beam current is given as
\begin{equation}
\label{Ie}
I_e= e N_e  f = \frac{P}{E_e},
\end{equation}
where $I_e$ is given in mA, $P$ is the electron beam power, in MW,
and $E_e$ the electron beam energy in GeV. From the values above one derives
that the current to reach $10^{33}$\,cm$^{-2}$s$^{-1}$ under the quoted conditions is 
$I_e=6.4$\,mA. This corresponds to $384$\,MW beam power at
$E_e=60$\,GeV. Given a $100$\,MW wall-plug power limit for the design
this can only be realized in an energy recovery 
mode. This implies CW operation which can be realized  with SC cavity
gradients of about $20$\,MV/m for two linacs of $1$\,km length each. 
The  configuration considered in the CDR uses
$P_0=24$\,MW linac grid power, which assumes an ERL efficiency of
$\eta=0.94$ and $P=P_0/(1-\eta)$. 
A total of $78$\,MW is foreseen assuming a cryogenics
power consumption of $21$\,MW, which may be reduced
with a quality factor $Q_0$ of the superconducting (SC)
cavities exceeding the assumed $2.5 \cdot 10^{10}$,
and $23$\,MW for the compensation of synchrotron losses 
in the return arcs. The quality of the SC cavity and mastering the
ERL technique are critical to the success of the LHeC.

The luminosity may be further enhanced 
because the proton beam brightness, $N_p/\epsilon_p$,
is expected to be larger by a factor of $2.5$ than here assumed,
the electron current may be doubled based on an enlarged
$Q_0$ value and $\beta^*$ could be reduced
to $5$\,cm. If all these improvements were realized the LHeC
would be an $ep$ collider with  a luminosity
of $10^{34}$\,cm$^{-2}$s$^{-1}$ enhancing substantially its
Higgs and BSM physics potential. Small corrections to Eq.\,\ref{LLR}
as are discussed in the CDR,
may be an hourglass reduction factor of $0.9$,
a luminosity enlargement for $e^-p$
from pinch effects of $1.35$, and perhaps
a reduction to $2/3$ if a clearing gap was introduced 
for fast ion stability. 
Table \ref{tab1} presents LHeC parameters, including, in parentheses,  
values for the increased luminosity version. 

\begin{table*}[htb]
%
\begin{center}
{\begin{tabular}{|l|cc|}
\hline
parameter [unit] &  \multicolumn{2}{|c|}{LHeC} \\ 
\hline
species & $e^{-}$ & $p$, $^{208}$Pb$^{82+}$ \\    
beam energy (/nucleon) [GeV] & 60 & 7000, 2760 \\
bunch spacing [ns] & 25, 100 & 25, 100 \\
bunch intensity (nucleon) [$10^{10}$]  & 0.1 (0.2), 0.4 & 17 (22), 2.5  \\
beam current [mA] & 6.4 (12.8) & 860 (1110), 6 \\
rms bunch length [mm] &  0.6 & 75.5  \\
polarization [\%] & 90 & none, none \\  
normalized rms emittance [$\mu$m]  & 50  & 3.75 (2.0), 1.5  \\
geometric rms emittance [nm] & 0.43  & 0.50 (0.31)  \\
IP beta function $\beta_{x,y}^{\ast}$ [m] & 0.12 (0.032) & 0.1 (0.05) \\   
IP spot size [$\mu$m]  & 7.2 (3.7) & 7.2 (3.7)\\   
synchrotron tune $Q_{s}$ &  --- & $1.9\times 10^{-3}$ \\ 
hadron beam-beam parameter &   \multicolumn{2}{|c|}{0.0001 (0.0002)} \\ 
lepton disruption parameter $D$  & \multicolumn{2}{|c|}{6 (30)} \\ 
crossing angle &   \multicolumn{2}{|c|}{0 (detector-integrated dipole)} \\ 
hourglass reduction factor $H_{hg}$  & \multicolumn{2}{|c|}{0.91 (0.67)} \\ 
pinch enhancement factor $H_{D}$  & \multicolumn{2}{|c|}{1.35} \\ 
CM energy [TeV] &  \multicolumn{2}{|c|}{1300, 810} \\ 
luminosity / nucleon [$10^{33}$ cm$^{-2}$s$^{-1}$] &  \multicolumn{2}{|c|}{1 (10), 0.2} \\ 
\hline
\end{tabular} \label{tab1}}
\end{center}
\caption{\footnotesize{LHeC $ep$ and $eA$ collider parameters. The numbers 
give the default CDR values, with
optimum values for maximum $ep$ luminosity in parentheses and values for
the $ePb$ configuration separated by a comma.}}
\end{table*}
\subsection{Components and frequency choice}
In the CDR~\cite{AbelleiraFernandez:2012cc},
designs of the magnets, RF, cryogenic and further components 
have been considered in quite some detail. Main parameters for both the
$RR$ and the $LR$ configurations are summarized in Tab.\,\ref{tabcomp}.
The total number of magnets (dipoles and quadrupoles excluding
the few special IR magnets) and 
cavities is $4160$ for the ring and $5978$ for the linac case.
The majority are the $3080~(3504)$ normal conducting
dipole magnets of $5.4~(4)$\,m length for the ring (linac return arcs),
for which short model prototypes have been successfully built,
testing different magnet concepts at
BINP Novosibirsk and at CERN as is described in the CDR. 
The number of high quality cavities for the two linacs is $960$,
possibly grouped in $120$ cavity-cryo modules. This
is an order of magnitude less than is required for the ILC.
For the RF frequency values significantly below $1$\,GHz are  
suggested by beam dynamics studies, RF power considerations
with $NbTi$ grain and operating temperature effects
and synchrotron loss compensation systems.   The specific value
has to be a multiple of the LHC bunch frequency and was
recently chosen to be $802$\,MHz for
genuine synergy with the HL-LHC higher harmonic RF system.
The cryogenics system 
for the linac critically depends on the cooling power per cavity, which for
the draft design is assumed to be $32$\,W at a temperature of $2$\,K.  This
leads to a cryogenics system with a total electric grid power of $21$\,MW.
The development of a cavity-cryo module for the LHeC
is directed to achieve a high $Q_0$ value and to reduce the
dissipated heat per cavity, which will reduce the dimension of
the cryogenics system.

\begin{table}[hbt]
\begin{center}  
   \begin{tabular}{|l|c|c|}
       \hline
       &  Ring &  Linac \\ 
       \hline
      magnets  & & \\
\hline
number of dipoles  &  $ 3080 $ & $ 3504 $   \\ 
dipole field [T] & $0.013-0.076$ &  $0.046-0.264$ \\
number  of quadrupoles  &  $ 968 $ & $ 1514 $   \\ 
\hline
      RF and cryogenics & & \\
\hline
number of cavities & $112$ &  $960$ \\
gradient [MV/m]  & $11.9$ &  $20$ \\
linac grid power [MW] & $ - $ & $24$ \\ 
synchrotron loss compensation [MW] & $49$ & $23$ \\ 
cavity voltage  [MV]  & $5$ & $20.8$ \\  
cavity $R/Q$ [$\Omega$] & 114  & $285$ \\ 
cavity $Q_0$ & $-$ & $2.5~10^{10}$ \\ 
cooling power [kW]  & $5.4$@$4.2$ K & $30$@$2$ K  \\ 
\hline
   \end{tabular}
   \end{center}
    \label{tabcomp}
\caption{\footnotesize{Selected components and parameters of
 electron  ring (left) and linac (right) accelerators, taken from the LHeC CDR.}}
\end{table}
\subsection{Further accelerator developments}
Following the publication of the CDR~\cite{AbelleiraFernandez:2012cc}
essential tests are now being prepared for various key components of the LHeC:
\begin{itemize}
\item
Superconducting RF technology for the development of cavities with high $Q_0$ in CW operation; 
\item
Superconducting magnet technology for the development of $Nb_3Sn$ magnets for quadrupole designs with mirror cross sections with apertures for high as well as low magnetic field configurations. This concerns specifically the prototyping of the $Q_1$ three-beam magnet nearest to the IP~\cite{AbelleiraFernandez:2012cc};
\item
Optimization for the design of normal conducting magnets suited 
for the return arcs of the energy recovery options with 
multiple magnet systems ($3$ per arc);
\item
Design of an LHeC ERL Test Facility (LTF)~\cite{tf}
 for studying and testing the various technical 
 components and building up operational 
 experience at CERN;
\item
Civil engineering studies for the Linac-Ring option including
the connection of the electron and proton tunnels;
\item
Design of the vacuum and beam pipe 
system for the experimental insertion of the LHeC.
\end{itemize}
Further studies are foreseen of the full optics 
and layout integration of the LHeC into the
high luminosity LHC project as well as of a suitable design to
maximize the positron intensity.
The goal of these developments is to prepare the ground for deciding
on the LHeC project later, in the context of the evolution of particle 
physics, in particular the LHC nominal beam energy results, and other projects
at CERN and beyond. As a new opportunity to further exploit the LHC at CERN, 
the LHeC  requires strong international efforts.
\subsection{Time schedule and mode of operation}  
The electron accelerator and new detector require a period of
about a decade to be realized, based on experience from previous
particle physics projects, as for example HERA, H1 and CMS. 
This duration fits with the industrialization and
production schedules, mainly determined by  
the required $\sim 3500$ approximately $5$\,m long warm
arc dipoles and  the $960$ cavities for the Linac.
The current lifetime estimates and physics plans for the LHC are for 
two more decades of operation, which currently points
to the shutdown LS3 for a major transition to this upgrade
in the mid twenties.

\section{Summary}
%

The LHeC is the natural, and the only possible 
successor of the DIS energy frontier exploration
in the coming decades.
It follows fixed target experiments at $\sim 10$\,GeV
and HERA at $\sim 100$\,GeV of cms energy 
in order to study  lepton-parton interactions  at $\sim 1000$\,GeV.
Its physics program has key topics ($WW \rightarrow Higgs$, 
RPV SUSY, $\alpha_s$, gluon mapping, PDFs, saturation, eAÉ)
which all are closely linked to the LHC (Higgs, searches as for 
lepto-quarks and SUSY particles at high masses, QGP ..). With 
an electron beam
upgrade of the LHC, the LHC can be transformed to a high precision
energy frontier facility which is crucial for understanding new and ``old" physics and  possibly for the long term sustainability of the LHC program too.

The LHeC will deliver vital information for future QCD developments (N$^3$LO,
resummation, factorisation,   non-standard partons, photon, neutron and 
nuclear structure, AdS/CFT, non-pQCD, SUSY..). As a giant
next step into DIS physics it promises to find new 
phenomena (gluon saturation, instantons, odderons,
and speculatively substructure of heavy, so far elementary particles).
A factor of almost $10^4$ increase in kinematic range 
for electron-ion scattering  leads to  accessing the range of saturation 
in the DIS region, where $\alpha_s$ is small, in both $ep$ and $eA$, 
to shed light on the QGP and the mysteries of hadronization in media and outside.

The default  electron beam configuration is a novel ERL (with less than  $100$\,MW 
wall-plug power demand) in racetrack shape which
is built toward the inside of the LHC ring and tangential to IP2. 
This is designed to deliver 
multi-$100$\,fb$^{-1}$ of luminosity, i.e. more than a hundred times
HERA's integrated luminosity.
    The LHeC is designed for synchronous operation with the LHC (three beams) 
    and should be operational for  the final decade of its lifetime. 
    This gives $10-12$ years for its realization.
 A detector concept is described in the CDR suitable for the Linac-Ring IR and to obtain full coverage and    ultimate precision. 

Half of the LHeC is operational. The other half requires next: an ERL test facility at CERN, 
IR related magnet and beam pipe prototype designs, to strengthen the LHC-LHeC physics links, to simulate and gradually to prepare for building the detector.  The LHeC has a most attractive and important program worth to be pursued which will also help
maintaining the diversity  collider particle physics needs to progress
in the exploration of nature. It also has the potential to become a next Higgs factory.

\section*{Acknowledgments}
This paper describes the effort of nearly $200$ physicists
in formulating the physics program, simulating major physics
channels, designing a suitable detector system, and 
considering in much detail the accelerator options for
a next and luminous electron-hadron collider. The authors
are grateful to all their colleagues on LHeC and also to the 
CERN directorate for continuing encouragement and support 
in difficult times. We further wish to acknowledge the support
of NuPECC and ECFA, the input
from the about $20$ project referees and especially the
risen interest in making the vision of continuing DIS real.
%

\end{document}